\documentclass[aps,12pt,superscriptaddress,amsfonts,amssymb,amsmath]{revtex4-2}
\pdfoutput=1

\usepackage{graphicx}
\usepackage{epsfig}
\usepackage{makeidx}
\usepackage{xcolor}
\usepackage{amsmath}
\usepackage{bm}
\usepackage[normalem]{ulem}

\begin{document}

\title{
Excited states of coherent harmonic qubits with long-range photon coupling and dissipation
} 

\author{L.\ Gamberale \footnote{Email address: luca.gamberale@mib.infn.it}}
\affiliation{Quantumatter Inc., New York 10023 NY, USA\\ LEDA srl, Universit\`a Milano Bicocca \\ I-20126 Milano, Italy}

\author{G.\ Modanese \footnote{Email address: giovanni.modanese@unibz.it}}
\affiliation{Free University of Bozen-Bolzano \\ Faculty of Engineering \\ I-39100 Bolzano, Italy}
\date{\today}

\linespread{0.9}

\begin{abstract}
It is known that ensembles of interacting oscillators or qubits can exhibit the phenomenon of quantum synchronization. In this work we consider a set of $N$ identical two-state systems that we call ``harmonic qubits'', because the kinetic part of their Hamiltonian is of the form $\omega_0 \sum_i a^\dagger_i a_i$, coupled through a multi-state ``photon'' mode subject to dissipation. It has been proven numerically that when the coupling between the qubits and the photon is sufficiently strong, the ensemble condenses into a ground state with negative energy, the energy gap is proportional to $N$ and there are clear cross correlations $\langle a^\dagger_i a_j \rangle$. Here we are interested into the energy spectrum of the excited states of this system. In order to obtain information on  the coherent transitions  we introduce a weak coupling of each qubit with an external oscillator of variable frequency $\omega$ and we check via Monte Carlo time evolution for which values of $\omega$ variations in the occupation of the external oscillator occur. After adding a second external oscillator coupled to the first only through the $N$ qubits, we also look at the energy transfer between the two external oscillators in dependence on their frequency, a transfer which is possible only through the excited states of the qubits.  
 Above threshold (when $E_0<0$) we find resonant transfer at frequencies which are definitely higher, and growing with $N$. This signals the presence of collective excited states, separated by large energy gaps, which are absent below threshold. 

\end{abstract}

\maketitle

\section{Introduction}

When $N$ quantum oscillators with the same frequency are coupled through a cavity photon mode \cite{faisal1987theory,brandes2005coherent,forn2019ultrastrong,tindall2020quantum,azzam2020ten}, above a certain coupling strength the system becomes coherent, with a negative ground state energy $E_0$ proportional to $N$. This simple model shows some analogies with the Dicke model \cite{dicke1954coherence,andreev1980collective,kirton2019introduction}, but also important differences. In the Dicke model, the kinetic terms of the single emitters are those of a spin-$\frac{1}{2}$ or two-level system, and the photon coupling terms become equivalent to a single coupling with the total spin. For the quantum oscillators the kinetic terms are of the form $a^\dagger_i a_i$ and the photon coupling terms remain independent; still, it is possible to demonstrate the condensation to a coherent ground state analytically in the limit of large $N$ \cite{gamberale2023coherent}, or numerically, for small $N$ \cite{gamberale2023numerical}, for ex.\ with \texttt{QuTiP} \cite{qutip1,qutip2}. Two crucial conditions must be satisfied in order to obtain condensation:

\begin{enumerate}
    \item oscillation amplitudes limited by some mechanism, allowing to avoid instability;
    \item coupling strength above a certain threshold (physically achievable or not, depending on the concrete situations).
\end{enumerate}

Concerning the first condition, in \cite{gamberale2023numerical} we have tested the model introducing an amplitude limitation mechanism implemented numerically with terms like $(a^\dagger_i)^4 a_i^4$ in the Hamiltonian. In \cite{gamberale2024spectral} we have considered an alternative and more physical stabilizing mechanism which restricts the number of excited states of the oscillators.

For the second condition, we have discussed in detail in \cite{gamberale2023coherent} how a 3D-enhancing factor in the coupling between  photons and oscillators allows to avoid the ``no-go'' theorem for photon condensation in a planar cavity \cite{andolina2019cavity} and reach the critical threshold. 

The next issue that we need to address in view of possible applications of the model is the structure of the coherent excited states. For reasons explained below, the knowledge of the energy spectrum is not sufficient, but we need information on the transition probabilities which takes into account the properties of coherence of the states involved. For this purpose we have developed an heuristic numerical recipe which essentially amounts to a simulation of the transitions (Sect.\ \ref{method}).

The energy spectrum of the system can actually be found directly for a sufficiently small number $N$ of oscillators (the dimension of the matrix operators increases exponentially with $N$). The plain spectrum, however, does not give in our opinion the required information on the relevant transitions. The ``small'' transitions in which only 1 or 2 oscillators change their states are present both under-threshold and above-threshold, making the spectrum very ``granular''. The question is, when in the coherent phase ``large" transitions begin to be important, i.e.\ transitions in which many oscillators change their state at the same time. 

It is useful, under this respect, to make a comparison with the results by Hagelstein and Chaudhary on the property of ``fractionation of a large quantum'' in the spin-boson model augmented with off-resonant loss \cite{cohen1973quantum,hagelstein2008level,hagelstein2008multiphoton}. This property is related, physically, to the coherent excitations that we consider here. A weak version of the fractionation effect is known to occur in the non-dissipative spin-boson model with Hamiltonian
\begin{equation}
    H_{spin-boson}=\Delta E \frac{S_z}{\hbar} + \hbar\omega_0 a^\dagger a + V \frac{2S_x}{\hbar}(a^\dagger + a)
\end{equation}
(where $S_x$, $S_z$ are spin operators, $a$, $a^\dagger$ are destruction/creation operators and $V$ is a coupling constant).

In this case, slow coherent transitions are found to occur, in which the large two-level system quantum $\Delta E$ is divided into a large number of quanta of the oscillator of small frequency $\omega_0$. Fractionation effects are greatly enhanced by the addition of dissipation, which prevents destructive interference among distant degenerate states. In spite of the simplicity of the spin-boson model, the exact or perturbative analysis of these effects is very difficult. In our model, there are many independent oscillators, making such analysis even more untractable, except possibly in the limit when $N$ is very large.

It is possible to increase the dimension of numerical simulations by using Monte Carlo (MC) solvers, which typically scale up only linearly in the dimension. MC solvers also allow to introduce a dissipation term into the dynamics. For physical applications of our model  it is important to include dissipation for the photon field, representing cavity losses and ohmic quenching. These dissipation phenomena are inevitable also when the photon field interacts with conductive media or metal micro-particles, like e.g.\ in the plasmonic Dicke effect \cite{bergman2003surface,pustovit2009cooperative,pustovit2010plasmon}.

The progression from our previous work in \cite{gamberale2023coherent,gamberale2023numerical} to the present calculation has required a technical effort that we felt was worth reporting on, for ourselves and for others who might try to replicate or extend our results. The general idea is to use the ready \texttt{QuTiP} package; this can clearly have some disadvantages, but also the advantage of a better portability and greater involvement of the community.

The rest of the paper is organized as follows. In Sect.\ \ref{method} we give the expression of the Hamiltonian of the system and then explain briefly our heuristic method for discovering the transitions to coherent excited states.
In Sect.\ \ref{test} we test the method, applying it to some simple systems with known spectrum. In Sect.\ \ref{results} we describe the results obtained for the full system of $N$ harmonic qubits with dissipation. Sect.\ \ref{conc} contains our conclusions. 
 Some technical details and examples of the \texttt{QuTiP} codes employed are reported in Sect.\ \ref{codes}.

\begin{figure}[h]
\centering
\includegraphics[width=.9\textwidth]{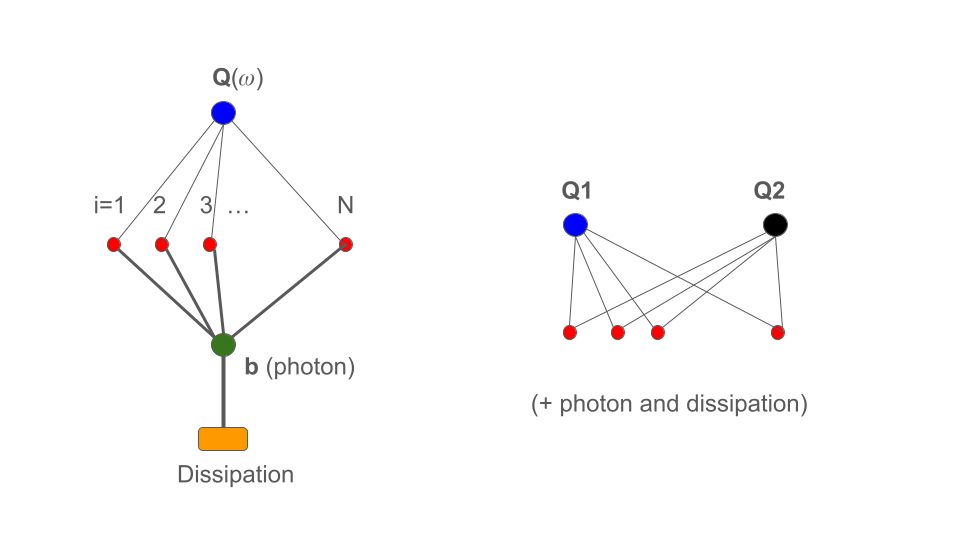}
\caption{General scheme of the considered quantum systems. On the \textbf{left}, the system described by the Hamiltonian of eqs.\ \eqref{H-tot} - \eqref{H-Q}, which comprises $N$ equal two-state harmonic oscillators (``harmonic qubits'') coupled quite strongly through a photonic mode with dissipation. Each oscillator is further weakly coupled to an external two-state oscillator $Q$ with variable frequency $\omega$. Upon varying $\omega$ one observes the occurrence, as expected, of ``Rabi'' oscillations when $\omega$ coincides with the characteristic frequency $\Delta E/\hbar$ of one of the transitions of the system. In a Rabi oscillation $\langle Q^\dagger Q \rangle$, the occupation number of $Q$,  drops significantly, see Figs.\ \ref{4ho-locale}, \ref{5ho-sotto-soglia}, \ref{5ho-sopra-soglia}, \ref{cfr-4HO-6HO-sopra-soglia}. Above a certain threshold strength of the coupling between the $N$ oscillators and the photon, the values of $\Delta E$ increase considerably, and grow with $N$. This signals the formation of collective synchronized states with large excitation energies. On the contrary, for small coupling strength the ``spectrum'' obtained displays only a broad maximum around the base frequency of the single oscillators (taken as $\omega_0=1$ in \eqref{H-0}). The \textbf{right} part of the figure represents the same system, to which a second external oscillator $Q_2$ has been added, coupled to the first oscillator indirectly via the weak coupling with the $N$ equal oscillators. For this system one obtains the double spectra of Figs.\ \ref{5ho-sotto-soglia}, \ref{5ho-sopra-soglia}, \ref{cfr-4HO-6HO-sopra-soglia}, starting from a state in which only $Q_1$ is excited. In Figs.\ \ref{5ho-sotto-soglia}, \ref{5ho-sopra-soglia}, \ref{cfr-4HO-6HO-sopra-soglia} the downward peaks show the decrease of the occupation of $Q_1$ at certain frequencies and the upward peaks represent corresponding increases in the occupation of $Q_2$. It is thus found that $Q_1$ and $Q_2$ can exchange energy at high frequency only when the $N$ oscillators are in a synchronized state thanks to a sufficiently strong coupling with the photon field.
}
\label{general-scheme}
\end{figure}

\section{Method and results}
\label{method}

The Hamiltonian of the considered system (see Fig.\ \ref{general-scheme}) with the base frequency $\omega_0=1$ is
\begin{equation}
    H=H_0+H_{int}+H_Q
\label{H-tot}
\end{equation}
\begin{equation}
    H_0=\sum_{i=1}^N a_i^\dagger a_i + b^\dagger b
\label{H-0}
\end{equation}
\begin{equation}
    H_{int}=\varepsilon \sum_{i=1}^N (a_i^\dagger + a_i)(b^\dagger + b)
\label{H-int}
\end{equation}
\begin{equation}
    H_Q=\omega Q^\dagger Q + \varepsilon_Q \sum_{i=1}^N (a_i^\dagger + a_i)( Q^\dagger +Q)
\label{H-Q}
\end{equation}
where $a_i$ are the destruction operators of the oscillators, $b$ that of the ``photon'' (of dimension $2N$) and $Q$ that of the external oscillator with variable frequency $\omega$ and dimension 2. The coupling $\varepsilon_Q$ is much smaller than $\varepsilon$. The $b$ oscillator is formally like the $a_i$ oscillators, apart from the larger dimension, but is coupled to all of them and in the possible applications is identified with a cavity photon mode, while the $a_i$ oscillators represent bound oscillating charges.

We recall that the diamagnetic term proportional to $\mathbf{A}^2$ is eliminated in Refs.\ \cite{gamberale2023coherent,gamberale2023numerical} through a suitable linear transformation on the photon field operators depending on the base frequency $\omega$ and on the renormalized frequency $\omega'$. In the present paper we are starting directly from the final form of the Hamiltonian obtained in \cite{gamberale2023coherent,gamberale2023numerical}. For this reason, the diamagnetic term is not present.

In order to find the excitation frequencies of the system, we use an heuristic method in which the frequency $\omega$ of the external oscillator $Q$ coupled to the $N$ oscillators $a_1,\ldots,a_n$ is varied and one observes for which values of $\omega$ the expectation value $\langle Q^\dagger Q \rangle$ of the occupation of the external oscillator drops significantly,  signaling a strong coupling effect similar to what happens in Rabi oscillations.  In most evolution runs the external oscillator starts from the first excited Fock state, while the $N$ oscillators $a_i$ and the photon field start from their Fock ground states. 

We refer to these oscillations as ``Rabi oscillations'' (in quotation marks) for brevity, as can be found in similar cases in the literature of open quantum systems (see e.g.\ \cite{manzano2020short}). In fact, we do not have here exactly what is usually meant in atomic physics by Rabi oscillations, whose frequency is defined by the coupling of a two-level system to an external electric field. Moreover, the oscillations we observe are strongly damped and for this reason other typical features of real Rabi oscillations, like e.g.\ the appearance of Mollow triplets, are not observed.

In \texttt{QuTiP}, the Monte Carlo solver is based on a quantum-jump approach to wave function evolution and allows for simulating an individual realization of the system dynamics. In this realization, the environment is continuously monitored, resulting in a series of quantum jumps in the system wave function, conditioned on the increase in information gained about the state of the system via the environmental measurements. In general, this evolution is governed by a Schr\"odinger equation with a non-Hermitian effective Hamiltonian which includes collapse operators.

In our simulations, the additional dissipation term is proportional to the photon destruction operator \texttt{sm}. In our previous work \cite{gamberale2023numerical} we have checked that the ground state energy does not depend on the proportionality coefficient. In this work we verified that the positions of the peaks in the $\omega$-spectrum are also independent on this coefficient.

\begin{figure}[h]
\centering
\includegraphics[width=.8\textwidth]{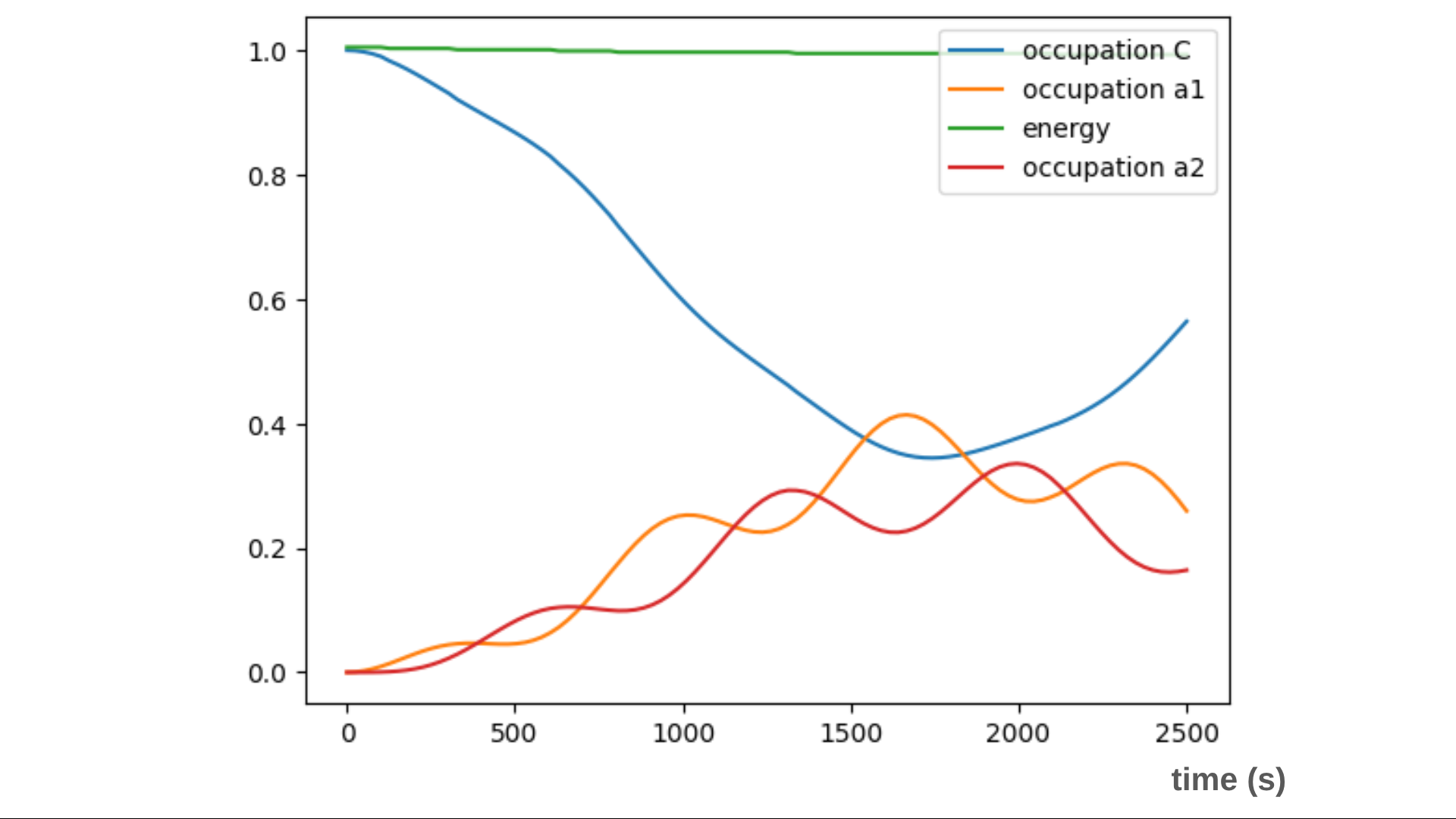}
\caption{Simple example of ``Rabi oscillations''.  Time evolution computed with a Monte Carlo algorithm of \texttt{QuTiP}, with weak dissipation, over a time interval (0,2500). The system comprises a harmonic oscillator $C$, initially in its excited state, weakly coupled to two oscillators $a_1$ and $a_2$ strongly coupled to each other. For a suitable value of the frequency of $C$, the occupation number $\langle C^\dagger C \rangle$ is seen to drop significantly. See code in Sect.\ \ref{codes}. 
}
\label{rabi-osc}
\end{figure}

\begin{figure}[h]
\centering
\includegraphics[width=0.9\textwidth]{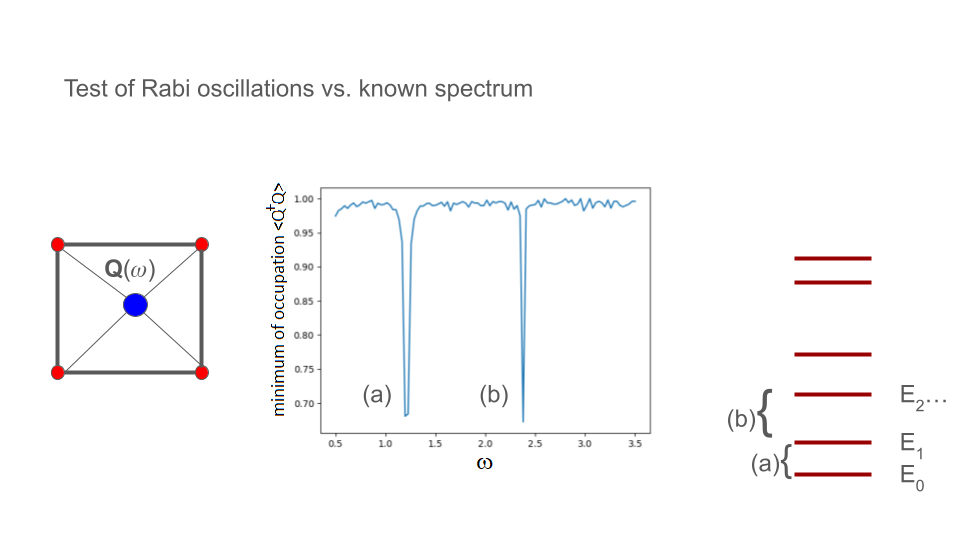}
\caption{Check of our indirect method for the observation of excitation energies, done in a case in which it is possible to compute directly the spectrum of the system. We have 4 oscillators strongly coupled to their neighbors and weakly coupled to an external oscillator $Q$ of variable frequency $\omega$. We let the system evolve in time in a similar way as shown in Fig.\ \ref{rabi-osc} for the simpler case of two oscillators, changing the values of $\omega$ and recording for each value the minimum reached by the occupation $\langle Q^\dagger Q \rangle$. A graph is obtained with peaks corresponding to the energy differences in the energy spectrum of the system. See complete graph and energy values in Fig.\ \ref{4ho-locale}.
}
\label{test-rabi}
\end{figure}

\subsection{Test with systems with known spectrum}
\label{test}

\begin{figure}[h]
\centering
\includegraphics[width=.45\textwidth]{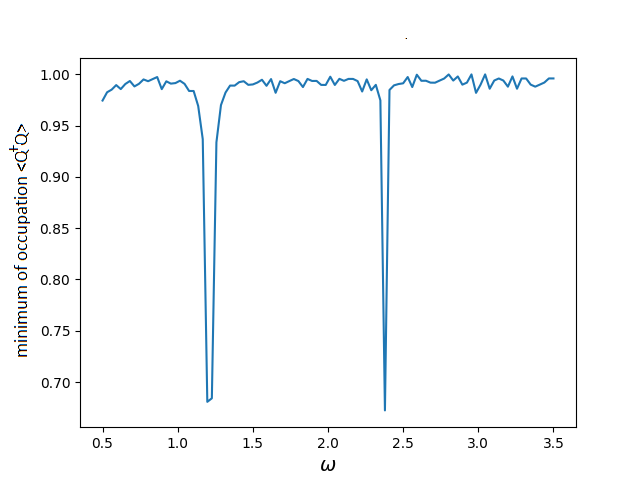}
\includegraphics[width=.45\textwidth]{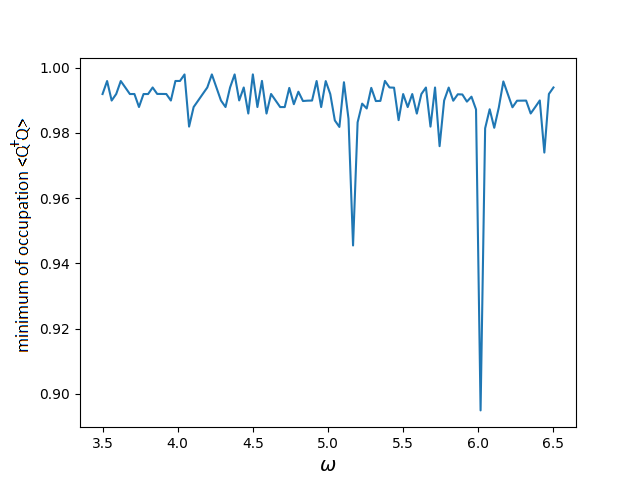}
\caption{Test runs of four oscillators with strong local coupling, but without photon coupling and dissipation (the system of Fig.\ \ref{test-rabi}). Large peaks at frequency 1.21, 2.37, 6.01 are visible. The complete energy spectrum is known to be [-4.17 -4.17 -4.16 -4.16  0.63  0.63  1. 1. 1. 1. 1.84 1.84 2. 2. 2. 2. 2. 2. 2. 2. 2.16 2.16 3. 3. 3. 3. 3.37 3.37 8.16 8.16 8.17 8.17].  (See Fig.\ \ref{spectrum}.) 
The peak at 1.21 is probably due to transitions between the states of energies (0.63,1.84) and (2.16,3.37); the peak at 2.37 corresponds to transitions (1.0,3.37) and that at 6.01 to transitions (-4.17,1.84), etc. Different peaks can be obtained by changing the initial states of the qubits and switching off the couplings between some qubits and the external oscillators.
}
\label{4ho-locale}
\end{figure}

This method can be tested with success on some simpler systems whose spectrum is known. For example, Fig.\ \ref{rabi-osc} shows the time evolution of the occupation of two weakly coupled quantum oscillators with close frequencies. Fig.\ \ref{test-rabi} shows the scheme of a system of 4 equal oscillators with reduced dimension $n=2$ coupled locally to each other and all weakly coupled to an external oscillator $Q$, thus with an Hamiltonian similar to \eqref{H-tot}, but with the photon field replaced by a local coupling to the neighbor oscillators. The spectrum of the system (without $Q$) can be easily computed, see caption of the figure. When the system is coupled to the external oscillator $Q$ and evolves in time starting from the ground states of the $a_i$ and the excited state of $Q$, we find oscillations in the occupation of $Q$ only when the frequency of $Q$ corresponds to some transitions. The code works as follows: after choosing a frequency interval containing approximately 100 to 200 values of $\omega$, for each value the system is evolved and the minimum value of $\langle Q^\dagger Q \rangle$ is recorded. All minimum values are plotted as a function of $\omega$, obtaining a sort of inverted spectrum (Fig.\ \ref{4ho-locale}) in which the descending peaks depend in general on the strength of the coupling to $Q$ (here fixed to a small value, typically $\varepsilon_Q \simeq 0.01$ ) and on the probability of the transition represented by the peak. This probability depends in turn on the degeneration/multiplicity of the involved eigenstates and on the way $Q$ is coupled to the $N$ oscillators. For ex., in Fig.\ \ref{test-rabi} if $Q$ is coupled only to 2 oscillators out of 4, the excitation spectrum changes, and the frequencies of the peaks correspond to transitions between different eigenstates, compared to the case when $Q$ is coupled to all 4 oscillators.

\begin{figure}[h]
\centering
\includegraphics[width=.8\textwidth]{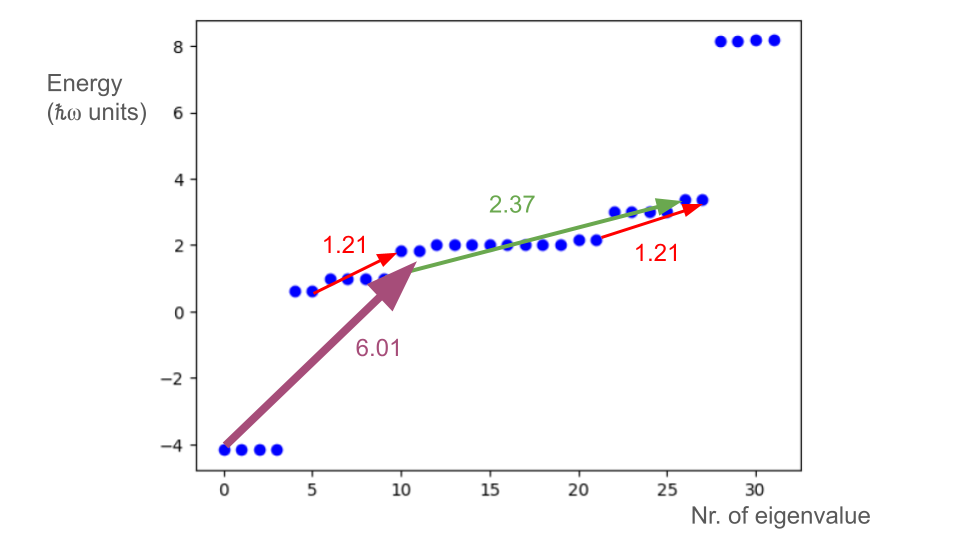}
\caption{ Graphical representation of the main transitions of the system in Fig.\ \ref{4ho-locale}. See further explanations in the text.  
}
\label{spectrum}
\end{figure}

Fig.\ \ref{spectrum} shows the energy spectrum, computed with \texttt{QuTiP}, of the simple benchmark system of Fig.\ \ref{4ho-locale} (four oscillators coupled strongly to their first neighbors and all weakly coupled to an external one). Each blue dot represents an energy eigenvalue. The $y$-axis gives the energy scale, the $x$-axis the progressive eigenvalue number. Several degenerate or almost-degenerate eigenvalues can be noticed; the exact energy values are listed in the caption of Fig.\ \ref{4ho-locale}. The arrows denote transitions whose energy differences correspond to peaks in the excitation probability of the external oscillator Q, i.e.\ peaks observed in simulation runs with a cycle over the frequency of Q. The colors of the arrows only help in relating each transition to its energy difference, written in the same color.

All this shows that the information on the excited states obtained with this method is in general not complete. The peaks observed, however, do represent real transition probabilities, including crucial coherence properties, which are very important for the system under consideration.  It is possible to plot the occupation of the levels as a function of time in correspondence to the frequencies of the peaks, confirming the existence of damped oscillations with frequency dependent on the coupling $\varepsilon_Q$. 

\subsection{Results for the system of qubits with dissipation}
\label{results}

A full theoretical justification and characterization of the proposed method clearly requires further elaboration. Here we just would like to show that it works and gives interesting results for our system, after the correct parameters of the MC evolution have been identified.

The choice of these parameters requires some care.
In fact, for each tested value of $\omega$ it is necessary that the system first reaches its coherent ground state in the presence of dissipation, and then, if there is a match between $\omega$ and the excitation frequencies, a Rabi oscillation begins, with a decrease of $\langle Q^\dagger Q \rangle$. If the dissipation term of the MC solver is too large, then the system reaches quickly its equilibrium but after that it is not able to resonate with $Q$. On the other hand, if dissipation is too small, then the system does not reach its ground state in a reasonable time, therefore its basis frequency is not stabilized, and again no resonance occurs.

\begin{figure}[h]
\centering
\includegraphics[width=.7\textwidth]{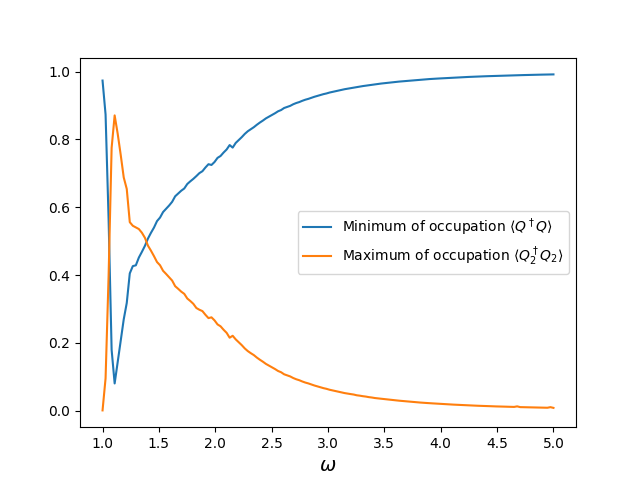}
\caption{Spectrum with $N=5$, $\varepsilon=0.2$ (5 qubits and photon coupling below threshold). The evolution time for each value of the frequency $\omega$ (150 values in total) is 4500 s. The couplings to the Q and Q2 oscillators are respectively 0.01 and 0.02. The total run takes about 15 days on a single CPU. It is clearly seen that the energy transfer between Q and Q2 only occurs at low frequency.
}
\label{5ho-sotto-soglia}
\end{figure}

\begin{figure}[h]
\centering
\includegraphics[width=.7\textwidth]{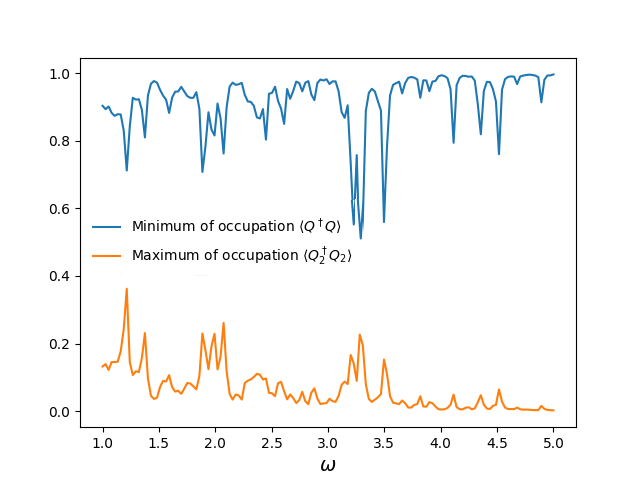}
\caption{Spectrum with $N=5$, $\varepsilon=0.7$ (5 qubits and photon coupling above threshold). The evolution time for each value of the frequency $\omega$ (150 values in total) is 4500 s. The couplings to the Q and Q2 oscillators are respectively 0.01 and 0.02.  The total run takes about 15 days on a single CPU. There is a clear increase in energy transfer between Q and Q2 at high frequency, compared to the below-threshold case.
}
\label{5ho-sopra-soglia}

\end{figure}

\begin{figure}[h]
\centering
\includegraphics[width=.6\textwidth]{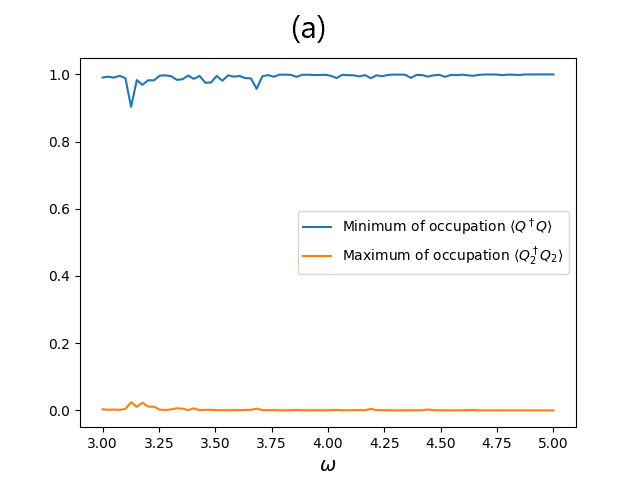}
\includegraphics[width=.6\textwidth]{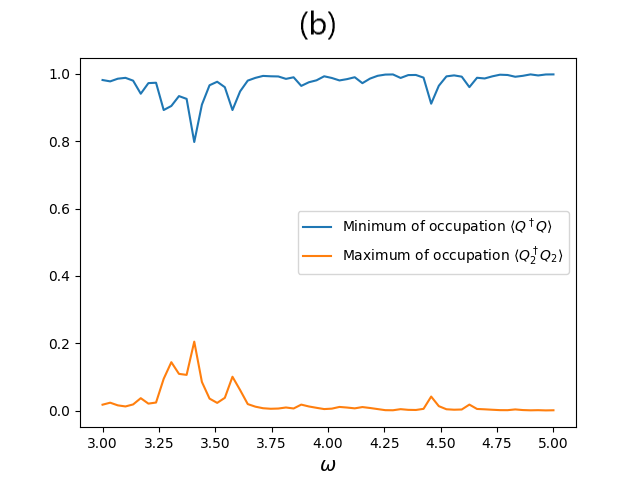}
\caption{(a) Spectrum of 4 qubits ($N=4$), above threshold, coupled with Q in state 1 and Q2 in state 0. (b) Same, with 6 qubits. There is a clear increase with $N$ in the energy transfer between Q and Q2 at high frequency.}
\label{cfr-4HO-6HO-sopra-soglia}
\end{figure}

Results are illustrated in Figs.\ \ref{5ho-sotto-soglia}, \ref{5ho-sopra-soglia} for the case of $N=5$ (5 qubits) and in Fig.\ \ref{cfr-4HO-6HO-sopra-soglia} (comparison between 4 and 6 qubits). Fig.\ \ref{5ho-sotto-soglia} shows the spectrum below threshold, i.e., for a value $\varepsilon$ of the coupling between qubits and photon such that the qubits do not reach a coherent/synchronized state. It is clearly observed that the energy transfer between the external oscillators Q and Q2 only occurs in a frequency interval about the base frequency of the qubits. In contrast, in Fig.\ \ref{5ho-sotto-soglia}, which represents an above-threshold case, several peaks at higher frequencies are visible, corresponding to excited states in which the qubits are ``synchronized'' \cite{zhirov2009quantum,lee2014entanglement,huan2020synchronization,vaidya2024exploring}. Finally, Fig.\ \ref{cfr-4HO-6HO-sopra-soglia} shows the effects on the energy transfer spectrum of an increase in the number $N$ of qubits. When $N$ is increased, the transition probabilities and the frequencies of the peaks also increase.

\section{Discussion and Conclusions}
\label{conc}

Comparing our results to those of Byrnes et al.\ \cite{byrnes2012macroscopic}, we note that in our case, constructive interference is observed as a function of the number $N$ of qubits, while the qubits always have two states only. On the contrary, in \cite{byrnes2012macroscopic} the number of qubits is fixed to 2, and $N$ represents the number of states of each qubit. 

In other studies concerning many-particle bosonic fields and coherence, a diffusion Monte Carlo method was applied to find the ground state properties of interacting charged bosons in one dimension confined in a harmonic double-well trap (\cite{jing2007ground} and refs.).

Compared to the present work, the calculation in \cite{jing2007ground} involves many more degrees of freedom, because the system comprises 64 identical bosons, each one described by a position coordinate $x_i$, while in our case we have at most 6 oscillators with 2 discrete states. The e.m.\ interaction of the bosons, however, is given by a classical Coulomb potential, while in our system the interaction among oscillators is mediated by a quantized radiation field.

We believe that the damping observed in the results of our simulations is a combination of the damping term inserted in the Monte Carlo solver (the term \texttt{[const. * sm]} which is the fourth argument of the \texttt{mcsolve} command, see \texttt{QuTiP} script), representing the ohmic quench and boundary effects, plus a dynamical damping which is a consequence of including multiple states and non-resonant excitations. In fact, the numerical quantum evolution performed by the solver is exact in the sense that it takes all states into account (then of course there can be a loss of accuracy due to numerical approximations). 

Comparing our system to the Jaynes-Cummings model we note two major differences. First, the Hamiltonian is somewhat different, involving for the qubits a kinetic term of the form $a^\dagger a$, as compared to one proportional to a Pauli matrix. Second, we did not investigate the dependence of the spectrum on the number of photon states; in each simulation, that number was kept fixed and equal to twice the number of qubits. Therefore we were not able to display the Jaynes-Cummings ``ladder" behavior proportional to the square root of the number of photon states.

In conclusion, we have found that passing from the calculation of the ground state in the coherent phase to that of the excited states is technically demanding. For this purpose we decided to make simulations of temporal evolution in which the system is coupled to external oscillators having a pre-defined frequency $\omega$. In each simulation, the system must first go into its coherent ground state (above-threshold). The relevant values of $\omega$ are not known in advance, therefore we need to make very long cycles in which $\omega$ is increased in small steps and the entire evolution (including condensation to the coherent ground state) is repeated. Each Monte Carlo evolution lasts several minutes, but the entire cycle takes a long time on a single CPU. Clearly after this ``explorative" work one should pass to the implementation of a parallel code quicker and more efficient, addressing explicitely precision issues. Until now we could not focus on this aspect because we needed to study the dynamics of the system across the parameter space and so we were just happy to use a ready-made solver.

The results of the previous sections show that when the system is in the coherent ground state, the excitation to states of higher energy involves a phenomenon of constructive interference of the $N$ qubits. The largest transition probability is obtained when the qubits change their state in a coordinated way. It may be expected that this effect of constructive interference is more pronounced when $N$ is increased. However, the values of $N$ tested so far are quite small.

The model expressed by the Hamiltonian \eqref{H-tot}, whose mathematical form is quite general, has been previously applied to a system of identical physical charges oscillating in a periodic 3D lattice. A system of this kind can condense into a coherent ground state, thanks to the growth of a trapped photonic mode of the same frequency, only if the spatial density of the oscillating charges is sufficiently large. This is necessary in order that the (negative) interaction term compensates for the positive energy of the photons. The present numerical simulations at small values of $N$ thus make sense, physically, if one assumes that the system is confined in a small volume, with periodic boundary conditions for the field.

Work is in progress for increasing the computational performance and the value of $N$, as well as for clarifying possible applications of the model, and for giving a more satisfactory theoretical basis to the simulation method employed.

\bigskip

\noindent
\textbf{Acknowledgments} - This work was partially supported by the Free University of Bozen-Bolzano with the research project NMCSYS-TN2815. G.M.\ is a member of INdAM (Istituto Nazionale di Alta Matematica).

\bigskip

\section{Codes}
\label{codes}

In the numerical simulations of oscillators in \cite{gamberale2023numerical} the dimension $n_{max}$ of the finite Hilbert space of states employed in the matrix representation of the operators is determined via a ``saturation'' criterium. For ex., in \cite{gamberale2023numerical} it was found that a dimension $n_{max}\geq 35$ was approximately necessary in order to obtain values of the ground state energy independent from $n_{max}$. The total space is a tensor product with dimension $n_{max}^N\cdot n_{photon}$, with $n_{photon} \simeq N n_{max}$. This means that by simulating 4, 5 or more oscillators one quickly reaches the largest dimension that can be handled by standard numerical eigenvalues methods. That is a problem, especially if one wants to obtain information about the excited states of the system.

After several trials of MC time evolution similar to those described in \cite{gamberale2023numerical} for the evaluation of the ground state energy in the presence of dissipation, we have concluded that in order to analyse the excited states it is necessary to consider, at least in a first approach, a radically simplified model in which the ``harmonic oscillators'' only have 2 possible states. This is obtained in \texttt{QuTiP} by defining the annihilation operators $a_i$ as \texttt{destroy(n)} with $n=2$. The hope is that by optimizing the method the calculation can be extended to the number of possible excited states found in \cite{gamberale2024spectral}, typically 12 for some real systems.

This dimensional reduction is necessary because the method proposed requires to repeat the MC time evolution of the system many times (typically a few hundreds), varying the proper frequency of a coupled external oscillator until resonant Rabi oscillations are found, signalling the existence of excited states. Such long cycles over frequency values are feasible only if each MC run takes at most, say, 10 minutes to 1 hour. This means we can handle no more than 5 or 6 oscillators (plus the photon oscillator and the external oscillator), unless parallel algorithms are employed.

For comparison, it can be noticed that if one wants to compute the entire spectrum of the system with matrix methods, in spite of the drastic simplification $n=2$ it is impossible to go beyond 10 or 12 oscillators, for CPU/memory reasons. And still this spectrum would not be of real interest, because no dissipation is included.

\noindent
\textbf{Code for Fig.\ \ref{rabi-osc}:}

\begin{verbatim}
from qutip import *
import matplotlib.pyplot as plt

times = np.linspace(0.0, 2500.0, 100)

n = 2
N=2

psi0 = tensor(fock(N, 1), fock(n, 0),fock(n,0))

# sm is the external oscillator, with frequency w

a1 = tensor(qeye(N), destroy(n),qeye(n))
a2 = tensor(qeye(N), qeye(n),destroy(n))
sm = tensor(destroy(N), qeye(n),qeye(n))

epsilon1 = 0.001
w=1.005

H0 = 1.0*(a1.dag() * a1 + a2.dag() * a2) + w*sm.dag() * sm 
Hint = (sm + sm.dag()) * (a1.dag() + a1) + 4*(a2 + a2.dag()) * (a1.dag() + a1)

H1 = H0 + epsilon1 * Hint

expect_ops=[sm.dag()*sm, a1.dag()*a1,H1,a2.dag()*a2,a1.dag()*a2]

result=mcsolve(H1,psi0,times,[0.002 * sm],expect_ops,ntraj=500)

plt.plot(times,result.expect[0],label='occupation C') 
plt.plot(times,result.expect[1],label='occupation a1')
plt.plot(times,result.expect[2],label='energy')
plt.plot(times,result.expect[3],label='occupation a2')
plt.legend(loc='upper right')

\end{verbatim}

\noindent
\textbf{Code for Fig.\ \ref{cfr-4HO-6HO-sopra-soglia}:}

\begin{verbatim}

# 4 qubits and photon and Q and Q2

from qutip import *
from numpy import sqrt, pi, array, sin, cos, arange
import matplotlib.pyplot as plt
import numpy as np

get_ipython().run_line_magic('matplotlib', 'inline')

Noscillators = 4
n = 2
N = 8
epsilon1 = 0.7
epsilon_Q = 0.01

times = np.linspace(0.0, 1500.0, 100)

a1 = tensor(qeye(N), destroy(n),qeye(n), qeye(n),qeye(n),qeye(n),qeye(n))
a2 = tensor(qeye(N), qeye(n),destroy(n), qeye(n),qeye(n),qeye(n),qeye(n))
a3 = tensor(qeye(N), qeye(n),qeye(n), destroy(n),qeye(n),qeye(n),qeye(n))
a4 = tensor(qeye(N), qeye(n),qeye(n), qeye(n),destroy(n),qeye(n),qeye(n))

Q= tensor(qeye(N), qeye(n),qeye(n), qeye(n),qeye(n),destroy(n),qeye(n))
Q2= tensor(qeye(N), qeye(n),qeye(n), qeye(n),qeye(n),qeye(n),destroy(n))

sm = tensor(destroy(N), qeye(n),qeye(n), qeye(n),qeye(n),qeye(n),qeye(n))

H0 = a1.dag() * a1 + a2.dag() * a2 + a3.dag() * a3 + a4.dag()*a4 + sm.dag() * sm 

AA = a1.dag() - a1 + a2.dag()-a2 + a3.dag()-a3 +a4.dag()-a4

Hint = sqrt(8*np.pi/3)/sqrt(Noscillators) * 1j/2 * (sm + sm.dag()) * AA

HintQ = 1j/2 * (Q + Q.dag() + Q2 + Q2.dag()) * AA

expect_ops=[Q.dag()*Q, Q2.dag()*Q2]

psi0 = tensor(fock(N, 0), fock(n,1),fock(n,1),fock(n,1),fock(n,0),fock(n,1),fock(n,0))

list_occ_Q=[]
list_occ_Q2=[]
frequencies = np.linspace(3.5, 6.5 , 120)

for w in frequencies:

 HQ = w*Q.dag()*Q + w*Q2.dag()*Q2
 H1 = H0 + epsilon1 * Hint + HQ + epsilon_Q*HintQ
 result=mcsolve(H1,psi0,times,[0.05 * sm],expect_ops,ntraj=500)
 oscillazioneQ=min(result.expect[0])
 oscillazioneQ2=max(result.expect[1])
 print("w,Q,Q2",w,oscillazioneQ,oscillazioneQ2)
 list_occ_Q.append(oscillazioneQ)
 list_occ_Q2.append(oscillazioneQ2)

line1, = plt.plot(frequencies,list_occ_Q)
plt.xlabel("$\\omega$",fontsize=14)
plt.ylabel("Minimum of occupation $\\langle Q^\dagger Q \\rangle$")

line2, = plt.plot(frequencies,list_occ_Q2)
plt.xlabel("$\\omega$",fontsize=14)
plt.ylabel("Maximum of occupation $\\langle Q2^\dagger Q \\rangle$")

plt.show()

\end{verbatim}

\bibliography{excited-states} 
\bibliographystyle{ieeetr}

\end{document}